%% file: paper.tex
\documentclass[letterpaper,twocolumn,10pt]{article}
\usepackage{usenix}
\pdfoutput=1

\usepackage[hyphens]{url}
\usepackage{hyperref}
\usepackage{graphicx}
\usepackage{fixltx2e}
\usepackage{tabu,array}
\newcolumntype{C}[1]{>{\centering\arraybackslash}m{#1}}
\usepackage{color}

\usepackage[font=small]{subfig}
\usepackage{amssymb} 
\usepackage{cite}

\newcommand{\secmodule}{Broker}

\begin{document}
	
\date{}

\title{Personalized Security Indicators to Detect \\Application Phishing Attacks in Mobile Platforms}

\author{
{\rm Claudio Marforio, Ramya Jayaram Masti, Claudio Soriente, Kari Kostiainen, Srdjan \v{C}apkun}\\
Department of Computer Science, ETH Zurich \\
\{firstname.lastname\}@inf.ethz.ch
} 

\maketitle

\subsection*{Abstract}

Phishing in mobile applications is a relevant threat with successful attacks reported in the wild.
In such attacks, malicious mobile applications masquerade as legitimate ones to steal user credentials.
In this paper we categorize application phishing attacks in mobile platforms and possible countermeasures.
We show that personalized security indicators can help users to detect phishing attacks and have very little deployment cost.
Personalized security indicators, however, rely on the user alertness to detect phishing attacks.
Previous work in the context of website phishing has shown that users tend to ignore the absence of security indicators and fall victim of the attacker.
Consequently, the research community has deemed personalized security indicators as an ineffective phishing detection mechanism.

We evaluate personalized security indicators as a phishing detection solution in the context of mobile applications.
We conducted a large-scale user study where a significant amount of participants that used personalized security indicators
were able to detect phishing. All participants that did not use indicators could not detect the attack and entered their credentials to a phishing application.
We found the difference in the attack detection ratio to be statistically significant.
Personalized security indicators can, therefore, help phishing detection in mobile applications and their reputation as an anti-phishing mechanism should be reconsidered.

We also propose a novel protocol to setup personalized security indicators under a strong adversarial model and provide details on its performance and usability.

\input{intro}
\input{taxonomy}
\input{userstudy}
\input{setup}
\input{relatedwork}
\input{conclusion}

{\footnotesize \bibliographystyle{acm}
\bibliography{references}}
\section*{Appendix}
\input{appendix}

\end{document}

%% file: intro.tex

\section{Introduction}
\label{sec:intro}

Application phishing attacks in mobile platforms occur when malicious
applications mimic the user interface (UI) of legitimate applications to steal
user data. Phishing applications have been reported in the wild~\cite{droid09, securelist}
and more sophisticated attacks are described in ~\cite{felt11w2sp, zu-woot12}.

Phishing applications do not exploit system vulnerabilities.
They use standard system features and leverage the user's incapacity to
distinguish the legitimate application from a phishing one. Therefore,
signature-based detection techniques (like the ones used to detect malware) and
security mechanisms on the device (like application sandboxing) cannot counter application phishing attacks.

Online services use \emph{personalized security indicators} to aid the user in
distinguishing the legitimate website from a phishing one~\cite{boa,vanguard}.
The personalized security indicator (or ``indicator'' from now on) is an image that the user
chooses when he enrolls for the online service. After enrollment, the website
displays the indicator every time the user logs in. The indicator allows the
user to authenticate the website and he should enter his credentials only if
the website displays the correct indicator.

Mobile applications could also use indicators to mitigate application
phishing attacks~\cite{zu-woot12}. The user chooses the indicator when he
installs the application and must check that the application shows the correct
indicator at each login. The indicator is then stored by the application and the
mobile OS prevents other applications from reading it.

In this paper we start by categorizing application phishing attacks in mobile
platforms and possible countermeasures. We show that personalized indicators can
mitigate all the attacks we consider and have little deployment cost.
Personalized indicators, however, rely on the user to detect phishing attacks by checking the presence of the correct indicator.
Previous work in the context of websites has shown that users tend to ignore the absence of indicators~\cite{schechter07sp,lee-w2sp14}.
Consequently, the research community deems personalized indicators as an ineffective phishing detection
mechanism~\cite{bravolillo11sp,hong12cacm}.

Because no study has evaluated personalized indicators for mobile platforms, we conducted a large-scale user study
to verify if personalized indicators can help mitigating application phishing attacks.
Over one week, 221 participants used a banking application we developed to complete various
e-banking tasks. On the last day, the application simulated a phishing attack.
A significant amount of participants that used personalized indicators were able to detect phishing.
All participants that did not use indicators could not detect the attack and entered their credentials to the phishing application.
We found the difference in the attack detection ratio to be statistically significant.
Based on these results we conclude that personalized indicators can help phishing detection in mobile applications and their reputation as an
anti-phishing mechanism should be reconsidered.

We also look at the problem of setting up personalized indicators in mobile applications.
Previous work~\cite{tygar96woec,dhamija05soups,zu-woot12,boa,vanguard} relies on the ``trust on first use'' (TOFU) assumption and does not account for the possibility that the indicator itself is phished during its setup.
We consider a stronger attacker model where malicious software is present on the device when the user sets up the indicator.
We propose a novel protocol to set up indicators securely under such adversary presence.
We implemented a prototype of the protocol for the Android platform and evaluated its performance.
We also present the results of a small-scale user study with 30 participants on the usability of the setup protocol.

To summarize, in this paper we make the following contributions:

\begin{itemize}
	\item We analyze application phishing attacks in mobile platforms and
possible countermeasures.
We show that personalized indicators are a cost-effective means to mitigate application phishing attacks, assuming user alertness.
	\item We report the results of a large-scale user study on the effectiveness of personalized indicators as a phishing-detection mechanism.
Our results show that personalized indicators can significantly improve detection of phishing attacks in mobile applications.
	\item We design and implement a novel protocol to securely set up personalized indicators under a strong adversarial model. We have implemented a prototype and evaluated its usability with a
small-scale user study.
\end{itemize}

The rest of the paper is organized as follows. In
Section~\ref{sec:comparison} we discuss application phishing attacks and
countermeasures.
In Section~\ref{sec:userstudy} we describe our user study and report its results.
We describe the secure setup protocol, its implementation and evaluation in
Section~\ref{sec:setup}. Section~\ref{sec:related} reviews related work and Section~\ref{sec:conclusion} provides concluding remarks.


%% file: taxonomy.tex

\section{Phishing Attacks and Countermeasures}
\label{sec:comparison}

In this section we categorize known application phishing attacks in mobile platforms.
All attacks are effective on Android while one of them also works for iOS.
We also discuss possible countermeasures and analyze them with respect to security, usability and deployment aspects.
Table~\ref{tab:solutions} provides a summary of phishing attacks and countermeasures.

\subsection{Phishing Attacks}
\label{subsec:attacks}

\noindent\textbf{Similarity attack.} The phishing application has a name, icon, and
UI that are similar or identical to the legitimate application. The adversary
must induce the user to install the phishing application in place of the
legitimate one.
Successful similarity attacks have been reported for Android~\cite{droid09, securelist, digitaltrends} and iOS~\cite{macrumors}.

\noindent\textbf{Forwarding attack.} Another phishing technique is to exploit the
application forwarding functionality of Android~\cite{felt11w2sp}. A malicious
application prompts the user to share an event (e.g., a high-score in a game)
on a social network and shows a button to start the social network application.
When the user taps the button, the malicious application does not launch the
social network application, but rather displays a phishing screen. The phishing
screen asks the user to enter the credentials to access his account on the
social network. Application forwarding is a common feature of Android and,
therefore, forwarding attacks may be difficult for the user to detect.

\noindent\textbf{Background attack.} The phishing application waits in the background
and uses the \texttt{ActivityManager} of Android, or a
side-channel \cite{lin-ndss14}, to monitor other running applications. When the
user starts the legitimate application, the phishing application activates
itself to the foreground and displays a phishing screen~\cite{felt11w2sp}.

\noindent\textbf{Notification attack.} The attacker shows a fake notification and
asks the user to enter his credentials~\cite{zu-woot12}. The notification
window can be customized by the adversary to mimic the appearance of the legitimate
application.

\noindent\textbf{Floating attack.}
The attacker can leverage the Android feature that allows one application to draw an activity on
top of the application in the foreground. This feature is used by applications
to always keep a window in the foreground, for example, to display floating
sticky notes. A phishing application that has the
\texttt{SYSTEM\_ALERT\_WINDOW} permission can draw a transparent input field on
top of the password input field of the legitimate application. The UI of the
legitimate application remains visible to the user that has no means to detect
the overlaid input field. When the user taps on the password field to enter his
password, the focus is transferred to the phishing application which receives
the password input by the user. 

\begin{table*}[tbh]
  \centering
  \scalebox{.8}{
  {\tabulinesep=.6mm
    \setlength{\tabcolsep}{1.5mm}
	\begin{tabu}{l|cc|ccc}
    \cline{2-6}
    &\multicolumn{2}{c|}{\textbf{Marketplace Phishing Detection}}&\multicolumn{3}{c}{\textbf{On-device Phishing Prevention}} \\
    \cline{2-6}
    & \multicolumn{1}{C{2cm}}{\textbf{Name similarity}} & \multicolumn{1}{C{2cm}|}{\textbf{Visual similarity}} & \multicolumn{1}{C{2cm}}{\textbf{Limit multi-tasking}} & \multicolumn{1}{C{2cm}}{\textbf{Application name}} & \multicolumn{1}{C{2cm}}{\textbf{Personal indicator}} \\
    \hline

    \textbf{attacks} & & & & & \\ \cline{1-1}
    similarity attack   & \checkmark & --          & --         & --         & \checkmark \\
    forwarding attack   & --         & \checkmark  & --         & \checkmark & \checkmark \\
    background attack   & --         & \checkmark  & \checkmark & \checkmark & \checkmark \\
    notification attack & --         & --          & \checkmark & \checkmark & \checkmark \\
    floating attack     & --         & --          & \checkmark & \checkmark & \checkmark \\

    \hline
    \textbf{security} & & & & & \\ \cline{1-1}
    false positives/negatives  & \checkmark & \checkmark & --         & --         & -- \\
    reliance on user alertness & --         & --         & --         & \checkmark & \checkmark \\

    \hline
    \textbf{usability} & & & & & \\ \cline{1-1}
    user effort at installation          & -- & -- & -- &-- & \checkmark \\
    user effort at runtime               & -- & -- & -- & \checkmark & \checkmark \\
    restrictions on device functionality & -- & -- & \checkmark & \checkmark$^1$ & -- \\

    \hline
    \textbf{deployment} & & & & & \\ \cline{1-1}
    changes to application provider (e.g., bank)        & -- & -- & -- & -- & --  \\
    changes to marketplace & \checkmark & \checkmark & -- & -- & -- \\
    changes to mobile OS   & \checkmark$^2$ & \checkmark$^2$ & \checkmark & \checkmark & -- \\
    changes to application & -- & -- & -- & -- & \checkmark \\
    \hline

    \multicolumn{6}{l}{$^1$restriction to full-screen applications with constant user interaction (Android Immersive mode)} \\
    \multicolumn{6}{l}{$^2$to run a security check on sideloaded applications} \\
  \end{tabu}}}
\caption{Comparison of mechanisms to prevent application phishing attacks in mobile platforms.}
  \label{tab:solutions}
\end{table*}

\subsection{Phishing Countermeasures}

None of the attacks we discuss exploit OS vulnerabilities, but they rather use
standard Android features and APIs. Therefore, security mechanisms on the
device (e.g., sandboxing or permission-based access control) cannot prevent
such attacks. Similarly, security screening run by the marketplace operator
(e.g., the Google Bouncer system\footnote{http://googlemobile.blogspot.ch/2012/02/android-and-security.html})
may not detect phishing applications.
Security screening relies on signature-based malware detection techniques that
look for specific system calls patterns and requested permissions. Phishing
applications do not exhibit malicious code patterns, nor they require suspicious permissions.

Similar to website phishing, application phishing prevention requires tailored
security mechanisms. We now describe possible countermeasures and categorize them in terms of security, usability and deployment.

\noindent\textbf{Name similarity.} Marketplace operators can attempt to detect
similarity attacks by searching for applications with similar names or icons. Since
many legitimate applications have similar names or icons (for example, banking
applications for the same bank in different countries), this approach would
produce a significant amount of false positives. Detecting phishing
applications at the marketplace does not rely on the user alertness nor does it
change the user experience. Checking for phishing applications downloaded from
the web or from third-party marketplaces (sideloading) requires changes to the
mobile OS. In particular, the mobile OS should query the marketplace operator
for applications that have the same name or icon of the application being
installed.

\noindent\textbf{Visual similarity.} The marketplace operator can attempt to mitigate
background or forwarding attacks by searching for applications with similar UIs
and, in particular, similar login screens. Since many applications share a
similar login screen, detection based on similar UIs is likely to cause false
positives. If detection is based on matching UIs, phishing applications that
use a slightly modified version of the legitimate application UI, may go unnoticed. Finding
an effective tradeoff (a similarity threshold) may be a challenging task and
requires further investigation.

\noindent\textbf{Limit multi-tasking.} An approach to counter background or floating
attacks is to limit multi-tasking on the device. The legitimate application can
trigger a restricted mode of operation where no third-party applications can
activate to the foreground. Multi-tasking can be re-enabled once the user
explicitly terminates the application. Activation to the foreground can always
be allowed for system services, in order to receive phone calls or SMS
messages. This approach does not rely on the user alertness but it requires
changes to the OS (to toggle multi-tasking on and off) and hinders the user
experience. For example, the user cannot receive events from third-party
applications (e.g., social network notifications) while he is interacting with
an application that requested multi-tasking to be disabled.

\noindent\textbf{Application name.}
The mobile OS can show a status bar with the name
of the application in the foreground~\cite{selhorst10trust}.
Phishing detection with this approach is effective only if the user is alert and the phishing
application has a name and icon that are noticeably different from the ones of
the legitimate application. This technique cannot address name similarity attacks.
Furthermore, the status bar reduces the screen estate for applications that run in full-screen mode. An approach where the
status bar appears only when the user interacts with the application is only
practical for application with low interaction, such
as video players (Android \emph{Lean Back} mode). For applications that require
constant interaction, such as games (Android \emph{Immersive} mode),
forcing a visible status bar would hinder the user experience.

\noindent\textbf{Personalized indicator.} When the application is installed, the user
chooses an image from his photogallery. When the application asks the user for
his credentials, it displays the image chosen by the user at installation time.
An alert user can detect a phishing attack if the application asking for his
credentials does not show the correct image. The mobile OS
prevents other applications from reading the indicator of a particular application (through application-specific storage). This
countermeasure can also mitigate floating attacks. In particular, the
legitimate application can check if it is running in the foreground and remove
the image when it detects that it has lost focus\footnote{An application can
detect that it loses focus by overriding the method
\texttt{onWindowFocusChanged()}.}. Personal indicators can be easily
deployed as they do not require changes to the OS or to the marketplace.
However, they demand extra user effort when the application is installed
(because the user must choose the indicator) and used (because the user must
check that the application displays the correct indicator).

\noindent\textbf{Summary.} Our analysis is summarized in Table~\ref{tab:solutions}.
All the countermeasures we discuss incur trade-offs in attack prevention,
usability, functionality and deployment.
Personalized indicators can address all the attacks we consider and are easy to deploy as they do not
require changes on the device or at the marketplace.
Personalized indicators, however, rely on the user to detect
phishing attacks. Previous work~\cite{schechter07sp,lee-w2sp14} has shown that most users ignore the absence
of indicators when logging into an online banking service from a desktop web browser.
These results contributed to the reputation of personalized indicators as a weak mechanism to detect phishing.
Because personalized indicators were never tested in the context of mobile platforms, we
decided to verify their effectiveness as a detection mechanism for application phishing attacks.
We conducted a large-scale user study and report its results in the next section.


%% file: userstudy.tex

\section{User Study}
\label{sec:userstudy}

\begin{figure}[t]
  \centering
  \subfloat[] {%
    \includegraphics[width=.45\linewidth]{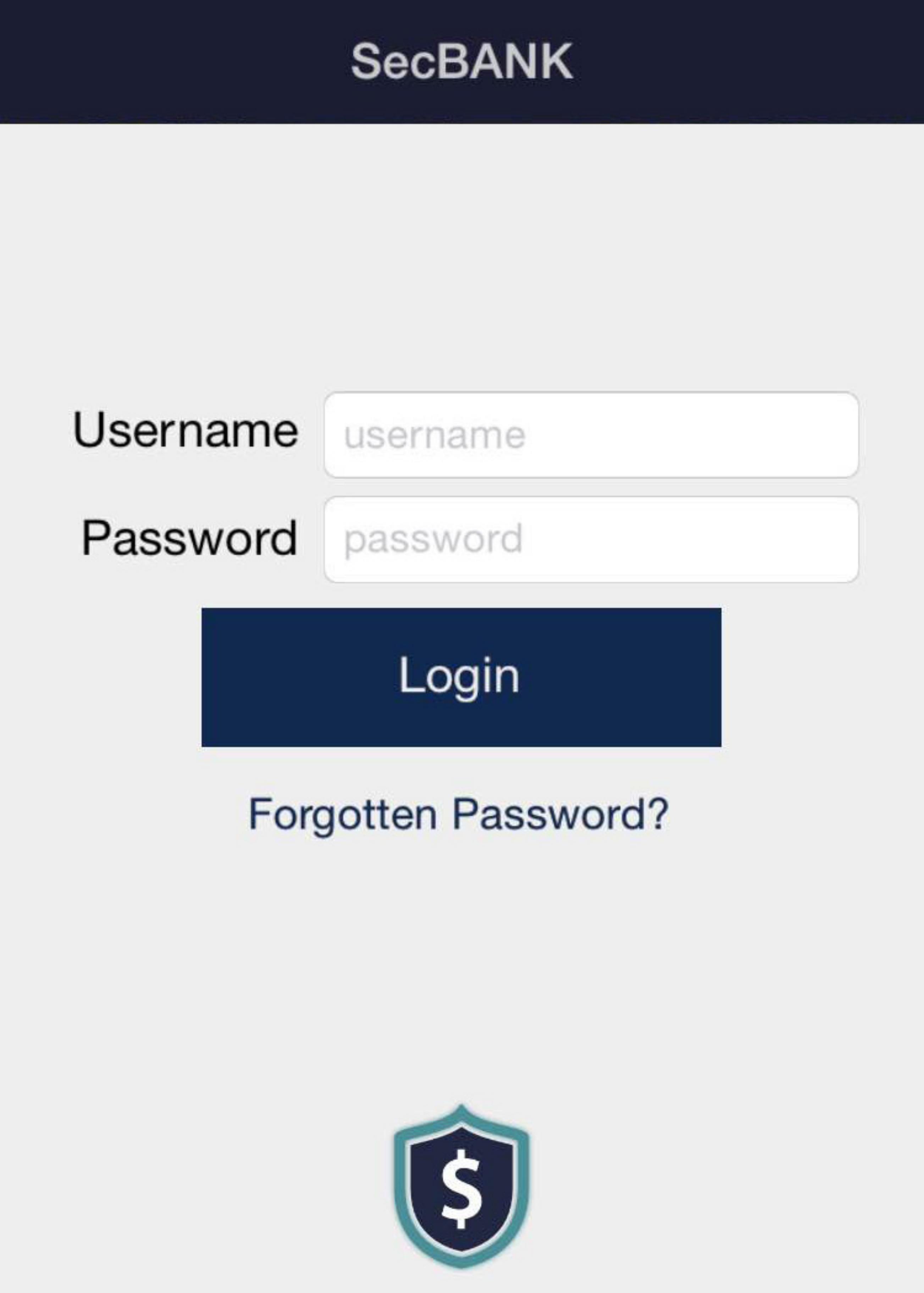}
    \label{app:control_group}
  }
  ~
  \quad
  \subfloat[]{%
    \includegraphics[width=.45\linewidth]{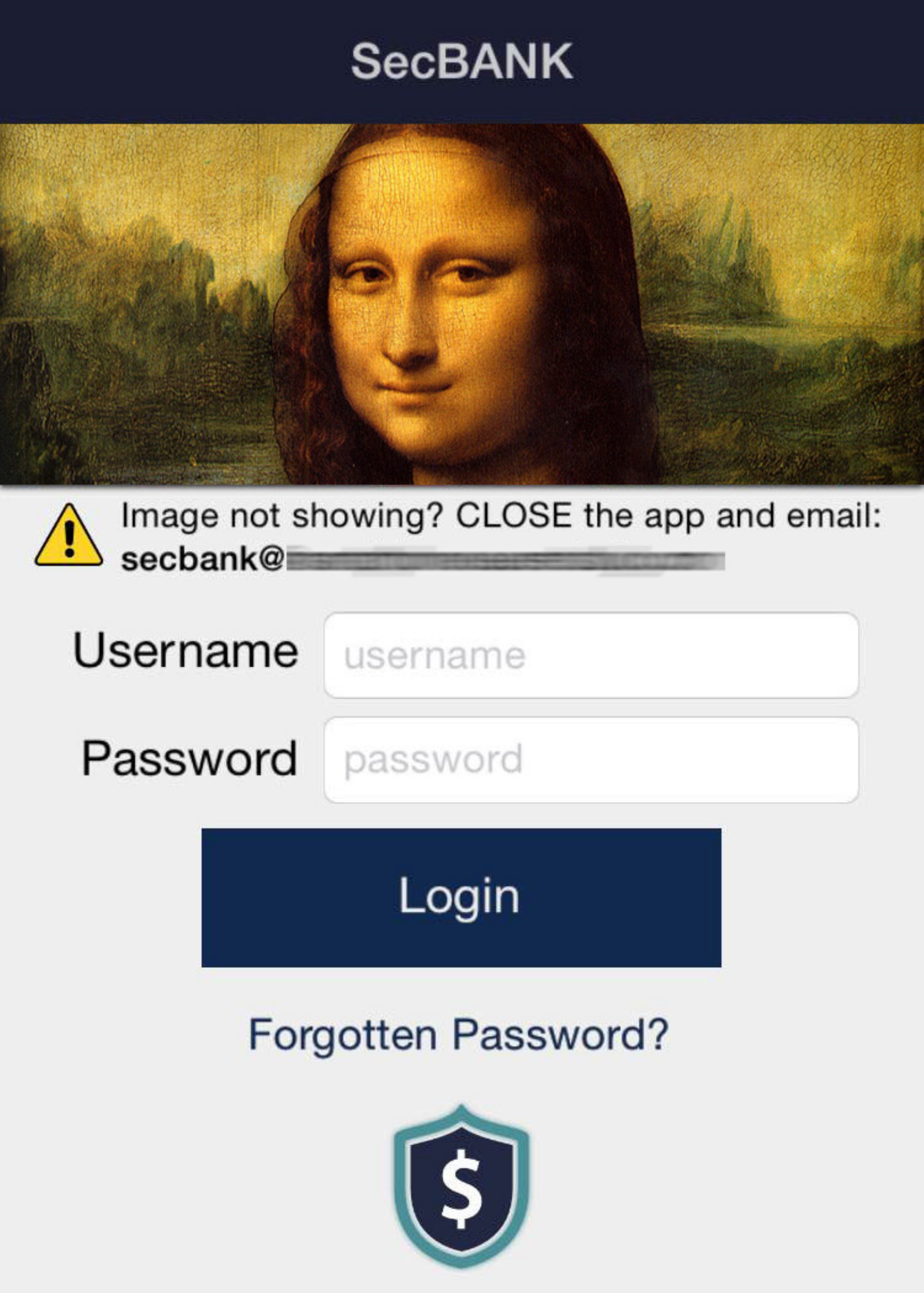}
    \label{app:attack_groups}
  }
  \caption{
  (a) SecBANK application for the control group. The application did not use personalized indicators.
  (b) SecBANK application for the three experimental groups. The application displayed the personalized indicators chosen by the user on the login screen. (In this example the user chose the Mona Lisa.) }
\end{figure}

The goal of our user study was to evaluate the effectiveness of personalized indicators as a phishing-detection mechanism for mobile applications. 
We focused on a mobile banking scenario and implemented an application that
allowed uses to carry out e-banking tasks for a fictional bank called SecBANK.
We asked participants to install the SecBANK application on their phones and
provided them with login credentials (username and password) to access their
accounts at SecBANK.
We assigned each participant to either a control group that used a SecBANK application without personalized indicators (Figure~\ref{app:control_group}), or one of
three experimental groups that used the SecBANK application with personalized indicators (Figure~\ref{app:attack_groups}).
The experimental groups differed in the type of phishing attack. The user study
lasted one week. During the first three days, participants carried out one
e-banking task per day to familiarize with the application and its login
mechanism (with or without personalized indicators, depending on the group to which a participant was assigned).
On the seventh day, participants were asked to perform a fourth e-banking task and, at this time,
the application simulated a background phishing attack.
We recorded whether or not participants entered their credentials while under attack.

We notified the ethical board of our institution which reviewed and approved our user-study protocol.


\subsection{Procedure}

\paragraph{Recruitment and Group Assignment.} We recruited participants through
an email sent to all people with an account at our institute (students, faculty
and university staff). The study was advertised as a user study to ``test the
usability of a mobile banking application'' without details on the real purpose
of our design. We offered a compensation of 20\$ to all participants who would
complete the pre-test questionnaire.

\begin{figure*}[t]
  \centering
  \subfloat[]{%
    \includegraphics[width=.22\linewidth]{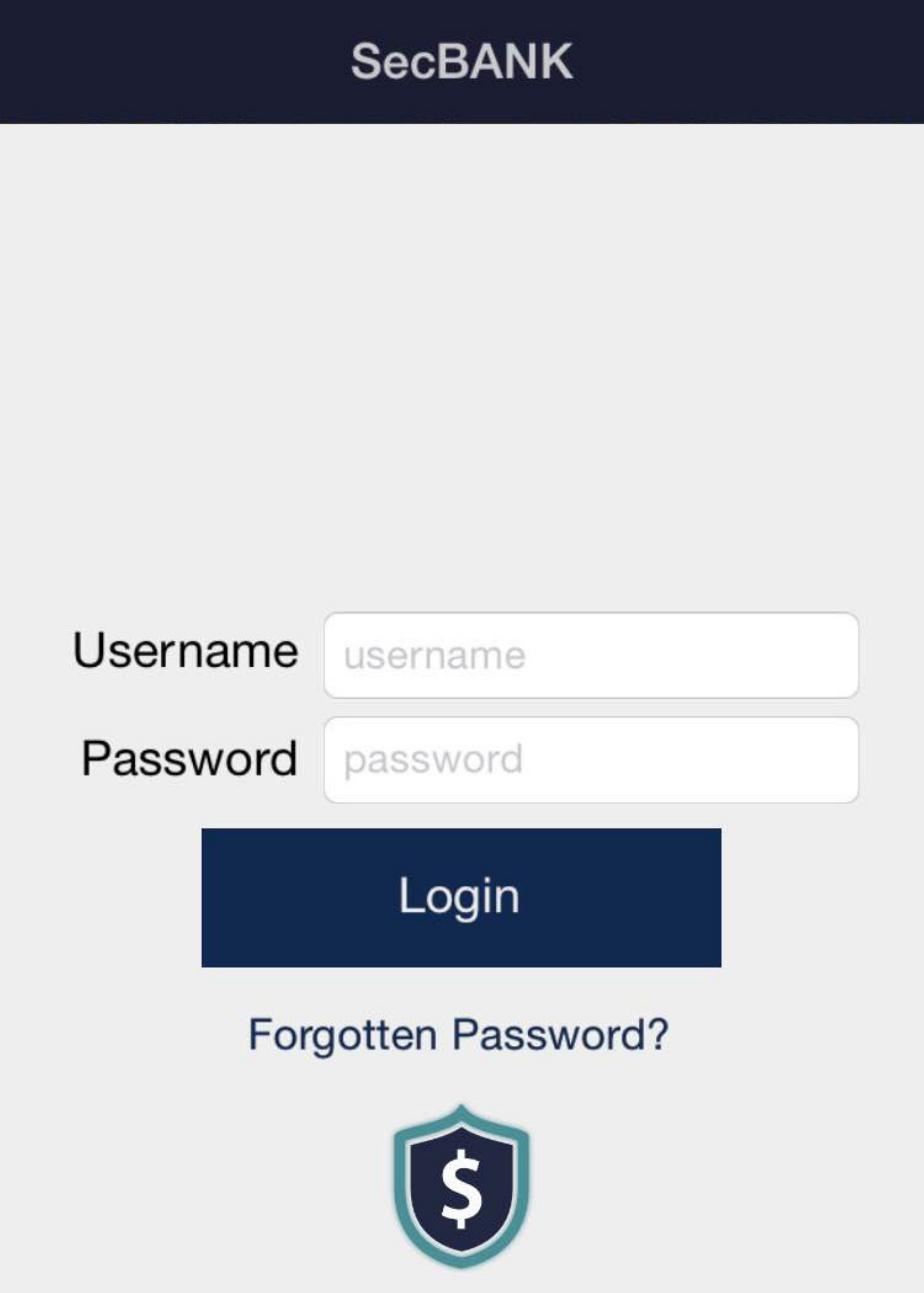}
    \label{app:noimage}
  }
  ~
  \quad
  \subfloat[]{%
    \includegraphics[width=.22\linewidth]{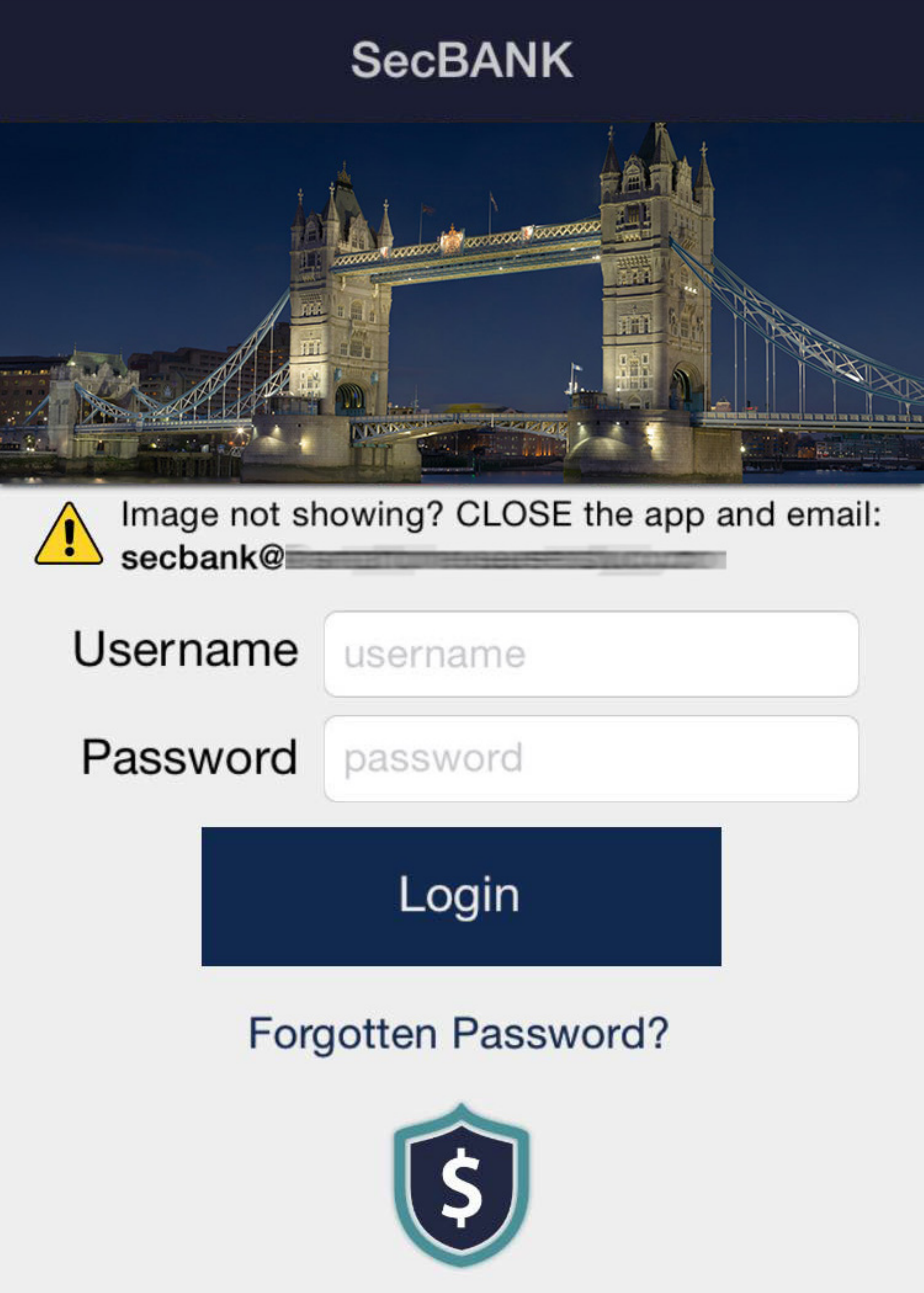}
    \label{app:towerbridge}
  }
  ~
  \quad
  \subfloat[]{%
    \includegraphics[width=.22\linewidth]{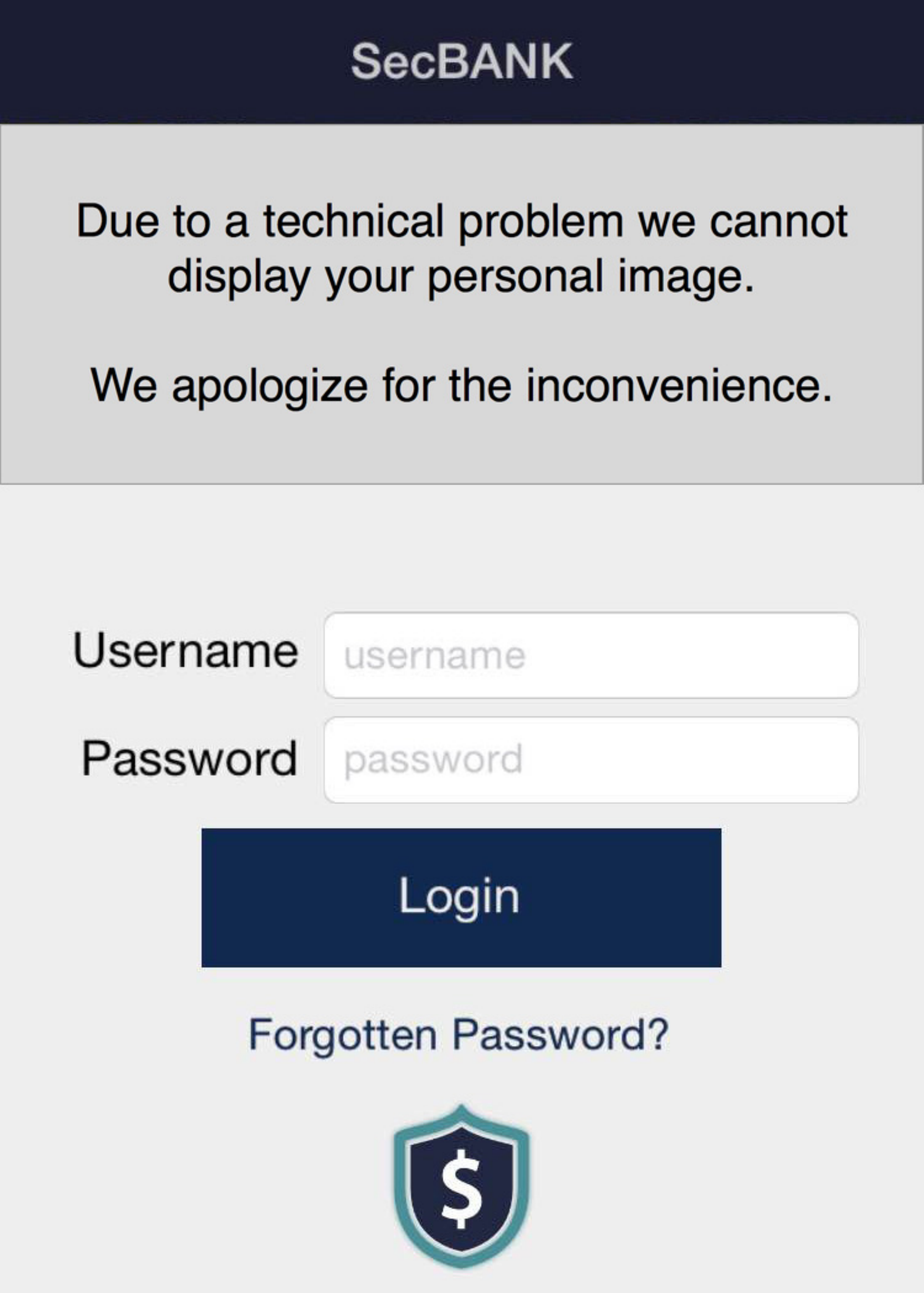}
    \label{app:maintenance}
  }
  \caption{
  (a) Missing-Image attack: The application does not show the indicator chosen by the user.
  (b) Random-Image attack: The application shows a random image from the local photogallery (in this case the Tower Bridge of London).
  (c) Maintenance attack: The application shows a message explaining that the indicator cannot be displayed due to technical reasons.
  }
\end{figure*}

We received 465 emails of potential participants to whom we replied with a link to an online pre-test
questionnaire designed to collect email addresses  and demographic information.
301 participants filled in the pre-test questionnaire. We assigned them to one of
the following four groups in a round-robin fashion:

\begin{itemize}
  \item \textbf{Control Group (A):} The application used by this group did not use personalized indicators.
  On the last day of the user study, the application simulated a background attack where the phishing application showed a clone of the SecBANK login screen.
  \item \textbf{Missing-Image Group (B):} The application used by this group supported personalized indicators. On the last day of the user study, the application simulated a background attack where the phishing application showed the SecBANK login screen without the indicator (Figure~\ref{app:noimage}).
  \item \textbf{Random-Image Group (C):} The application used by this group supported personalized indicators. On the last day of the user study, the application simulated a background attack where the phishing application showed the SecBANK login screen with a photo taken from the local photogallery that was different from the indicator chosen by the user (Figure~\ref{app:towerbridge}).
  \item \textbf{Maintenance Group (D):} The application used by this group supported personalized indicators. On the last day of the user study, the application simulated a background attack where the phishing application showed the SecBANK login with an ``under maintenance'' notification in place of the indicator chosen by the user (Figure~\ref{app:maintenance}).
\end{itemize}

We sent an email to all participants who completed the pre-test questionnaire
with a link to a webpage from which they could download and install the SecBANK
application. Participants in the Control Group (A) were directed to a webpage
where we only explained how to download and install the application.
Participants in experimental groups B, C, and D were directed to a webpage that
also explained the concept of personalized indicators. The
webpage advised that participants should not enter their login credentials if
the application were not showing the correct indicator.
The instructions were similar to the ones provided by~\cite{boa,vanguard} to the users of their e-banking platforms.
276 participants visited the webpages and installed the SecBANK application on their devices.

After the installation, the SecBANK application for groups B, C, and D showed explanatory overlays to guide participants in choosing a personalized indicator from their photogallery.

\paragraph{Tasks.} The study lasted one week. Participants where asked to
perform four e-banking tasks on days 1, 2, 3, and 7. We sent instructions to
complete each task via email and asked participants to complete the task within
24 hours. The tasks were the followings:

\begin{itemize}
	\item \textbf{Task 1} (Day 1): Transfer 200\$ to ``Anna Smith''.
	\item \textbf{Task 2} (Day 2): Download the bank statement from the ``Account Overview'' tab.
	\item \textbf{Task 3} (Day 3): Activate the credit card from the ``Cards'' tab.
	\item \textbf{Task 4} (Day 7): Transfer 100\$ to ``George White''.
\end{itemize}

The goal of tasks 1--3 was to help participants to become familiar with the SecBANK
application.

Instructions to perform task 4 were sent four days after (including a week-end) the completion of task 3.
During this last task, the application simulated a
background attack against participants of all groups.
Participants in the Control Group (A) saw a login screen that matched the one of their SecBANK application.
Participants in the Missing-Image Group (B) saw the SecBANK login screen
without any personalized indicator (see Figure~\ref{app:noimage}). Participants in
the Random-Image Group (C) saw the SecBANK login screen with a random image
from their photogallery (e.g., the Tower Bridge as shown in
Figure~\ref{app:towerbridge}). Finally, participants in the Maintenance Group (D) saw a
message explaining that due to technical problems the indicator could not be
displayed (see Figure~\ref{app:maintenance}).

\begin{table}[t]
    \begin{center}
        \begin{tabu}{|l|c|} \hline
          \multicolumn{2}{|l|}{Gender} \\ \hline
          Male & 150 (68\%) \\
          Female & 71 (32\%) \\ \hline
          \multicolumn{2}{|l|}{Age} \\ \hline
          Up to 20 & 43 (20\%) \\
          21 -- 30 & 164 (74\%) \\
          31 -- 40 & 9 (4\%) \\
          41 -- 50 & 3 (1\%) \\
          51 -- 60 & 0 (0.00\%) \\
          Over 60 & 2  (1\%) \\ \hline
          \multicolumn{2}{|l|}{Use smartphone to read emails} \\ \hline
          Yes & 214 (97\%) \\
          No & 7 (3\%) \\ \hline
          \multicolumn{2}{|l|}{Use smartphone for social networks} \\ \hline
          Yes & 218 (99\%) \\
          No & 3 (1\%) \\ \hline
          \multicolumn{2}{|l|}{Use smartphone for e-banking} \\ \hline
          Yes & 97 (44\%) \\
          No & 124 (56\%) \\ \hline
        \end{tabu} 
    \end{center}
    \caption{Demographic information of the 221 participants that completed all tasks.}
    \label{tab:demographics}
\end{table}

\subsection{Results}

Out of 276 participants that installed the application, 221 completed all
tasks. We provide their demographics and other information collected during the
pre-test questionnaire in Table~\ref{tab:demographics}. The majority of the
participants were male (68\%) and thirty years old or younger (94\%). Most participants
used their smartphone to read emails (97\%) and access social networks (99\%).
Slightly less than half of the participants (44\%) used their smartphones for
mobile banking.

\begin{table}[t]
    \begin{center}
      {\tabulinesep=.7mm
        \setlength{\tabcolsep}{1.5mm}
        \begin{tabu}{|m{2.5cm}|C{2.2cm}|C{2.2cm}|} \hline
          & Phishing attack not detected & Phishing attack detected \\ \hline
          Control Group & 56 (100\%) & 0 (0\%) \\ \hline
          Missing-Image Group & 25 (45\%) & 30 (55\%) \\ \hline
          Random-Image Group & 33 (59\%) & 23 (41\%)\\ \hline
          Maintenance Group & 25 (46\%) & 29 (54\%) \\ \hline
        \end{tabu}}
    \end{center}
    \caption{Detection rate of the phishing attack in each group.}
    \label{tab:primary}
\end{table}

The 221 participants that completed all tasks were distributed as follows: 56
in the Control Group (A), 55 in the Missing-Image Group (B), 56 in the
Random-Image Group (C), and the remaining 54 in the Maintenance Group
(D)\footnote{We note that, by chance, the participants that completed all tasks were almost evenly distributed among the four groups.}.

\paragraph{Indicator Effectiveness.} Table~\ref{tab:primary} shows the
percentage of participants that detected the phishing attack during task 4.
None out of the 56 participants in the Control Group (A) detected the phishing
attack (i.e., all participants entered their login credentials to the phishing application). 
82 out of the 165 participants in groups B, C, and D detected the phishing
attack and did not enter their login credentials.

To analyze the statistical significance of these results we used the following
null hypothesis: ``there will be no difference in phishing detection between
users that use personalized indicators and users that do not use personalized indicators''. A chi-squared test showed that the difference was statistically
significant ($\chi^2(3,N = 221) = 46.96, p < 0.05$), thus the null hypothesis
can be rejected. We concluded that, in our user study, personalized indicators
improved the detection of application phishing attacks in mobile platforms.

\paragraph{Difference Between Attacks.} 
A closer look at the performance of participants in groups B, C, and D reveals that:
25 out of 55 participants in the Missing-Image Group (B),
33 out of 56 participants in the Random-Image Group (C), and
25 out of 54 participants in the Maintenance Group (D),
detected the phishing attack.

To analyze the detection rates of the different attack types we used the
following null hypothesis: ``participants will equally detect the three
phishing attacks we tested''. A chi-squared test showed no statistical
significance, thus we fail to reject the null hypothesis.
We concluded that in our test setup the three attacks were equally detected.

\paragraph{Other Factors.} We further analyzed the collected data to understand
if there were any correlation between the ability to detecting phishing and
gender, age, smartphone display size or familiarity with mobile banking, respectively. We
did not find any statistical significance for any of the factors we considered.
Table~\ref{tab:secondary} provides the result break-down.

Finally, we could not find any correlation between the ability to detect
phishing and the time participants spent setting up the personalized indicator or
logging in, respectively.
The mean time spent setting up the indicator for
participants that detected the phishing attack was 43s ($\pm$28s); the mean time for
participants that did not detect the phishing attack was 46s ($\pm$28s).

\begin{figure}[t]
  \centering
  \subfloat[Personal indicator setup] {%
    \includegraphics[width=.8\linewidth]{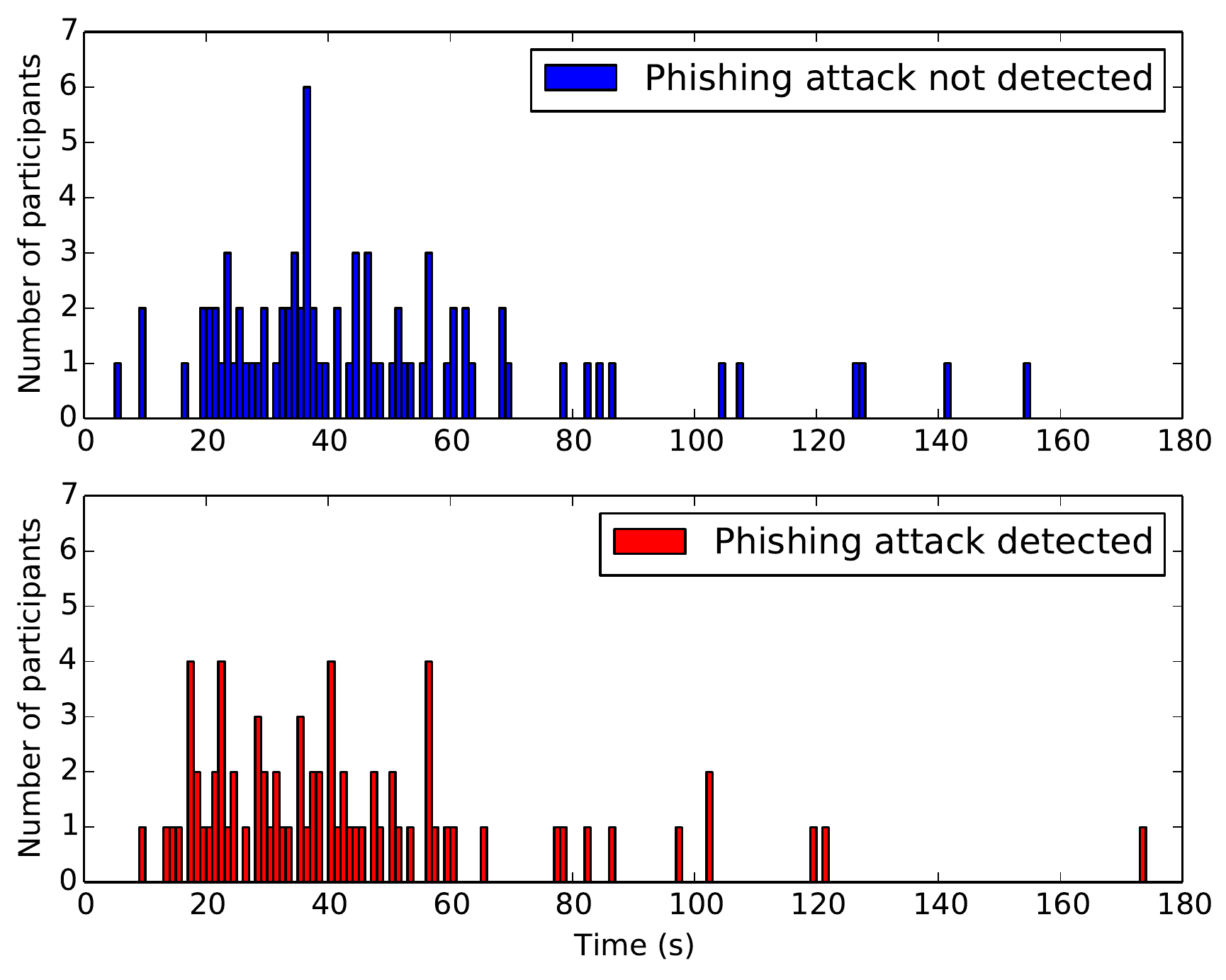}
    \label{fig:distindicator}
  } \\
  \subfloat[Login screen]{%
    \includegraphics[width=.8\linewidth]{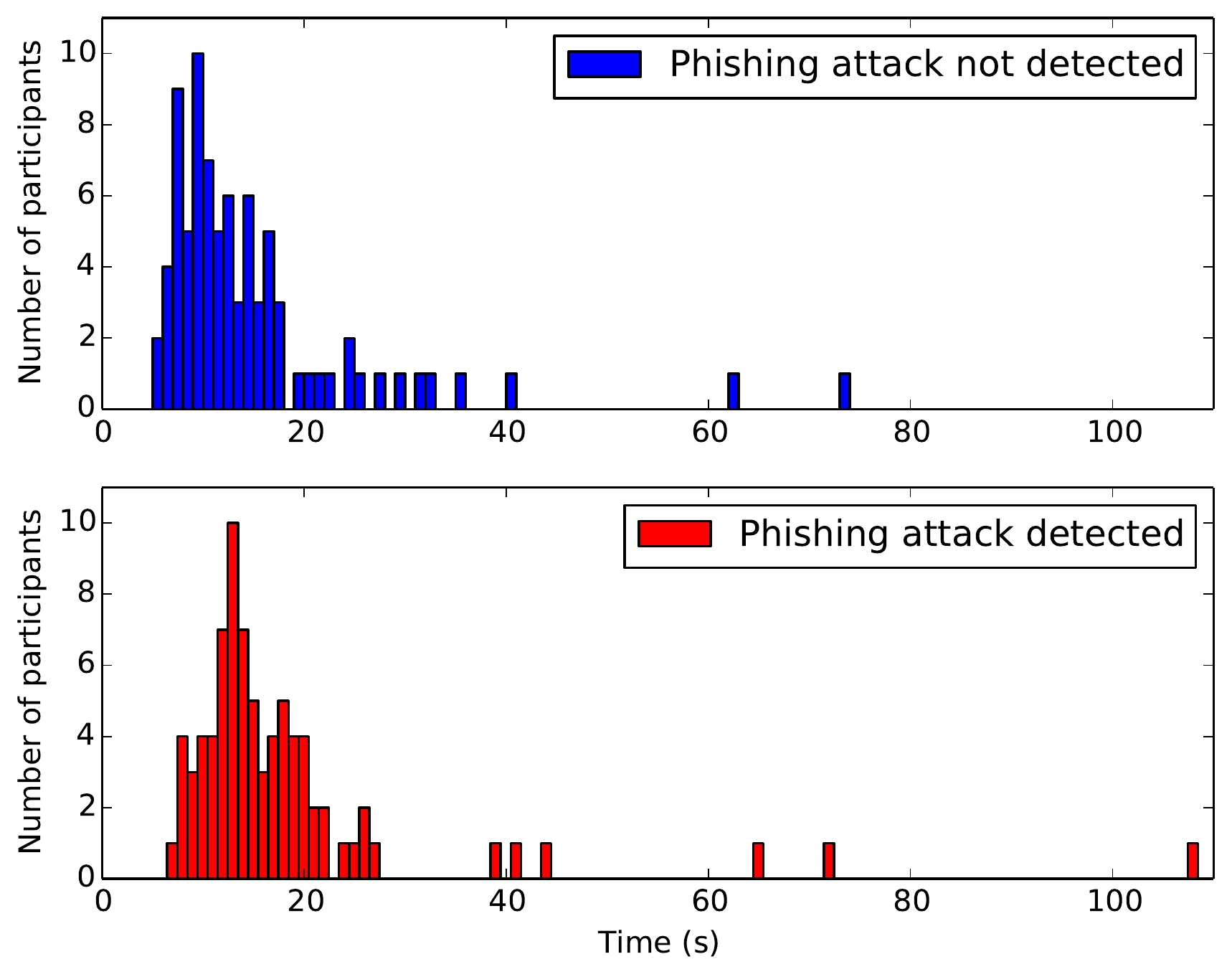}
    \label{fig:distlogin}
  }
  \caption{Distribution of the time spent by participants setting up the personalized indicator (a) and logging into the SecBANK application (b).}
\end{figure}

The mean time spent on the login screen for participants that detected the
phishing attack was 18s ($\pm$14s); the mean time for participants that did not
detect the phishing attack was 14s ($\pm$10s). The distribution of the times
spent while setting up the indicator and while logging in are shown in
Figure~\ref{fig:distindicator} and Figure~\ref{fig:distlogin}, respectively.

We could not find any correlation between the time spent setting up the indicator and the detection of a phishing attack.
Nonetheless, our results show that participants that detected the phishing attack
did not spend more time on the login screen than participants that did not detect phishing.
Our results show, therefore, that phishing attacks can be quickly detected by using personalized indicators.

\begin{table}[t]
    \begin{center}
      {\tabulinesep=.6mm
        \setlength{\tabcolsep}{1.5mm}
        \begin{tabu}{|m{2.5cm}|C{2.2cm}|C{2.2cm}|} \hline
          & Phishing attack not detected & Phishing attack detected \\ \hline
          \multicolumn{3}{|l|}{Gender} \\ \hline
          Male & 54 (48\%) & 59 (52\%) \\ \hline
          Female & 28 (56\%)& 23 (44\%) \\ \hline
          \multicolumn{3}{|l|}{Age} \\ \hline
          Up to 20 & 20 (57\%) & 15 (43\%) \\ \hline
          21 -- 30 & 61 (52\%) & 57 (48\%) \\ \hline
          31 -- 40 & 1 (14\%) & 6 (86\%) \\ \hline
          41 -- 50 & 1 (33\%) & 2 (67\%) \\ \hline
          51 -- 60 & 0 (0\%) & 0 (0\%) \\ \hline
          Over 60 & 0 (0\%) & 2 (100\%) \\ \hline
          \multicolumn{3}{|l|}{Use smartphone for e-banking} \\ \hline
          Yes & 35 (46\%) & 41 (54\%) \\ \hline
          No & 48 (54\%) & 41 (46\%) \\ \hline
          \multicolumn{3}{|l|}{Smartphone display size (diagonal)} \\ \hline
          up to $4$in & 20 (42\%) & 28 (58\%) \\ \hline
          from  $4$in to $4.5$in & 54 (55\%) & 44 (45\%) \\ \hline
          from $4.6$in to $5$in & 9 (47\%) & 10 (53\%) \\ \hline
        \end{tabu}}
    \end{center}
    \caption{Detection rate of the phishing attack in relation to gender, age, familiarity with mobile banking, and smartphone display size.}
    \label{tab:secondary}
\end{table}

\subsection{Post-test questionnaire}
\label{sec:posttest}

\begin{figure}[t]
	\centering
	\includegraphics[width=\columnwidth]{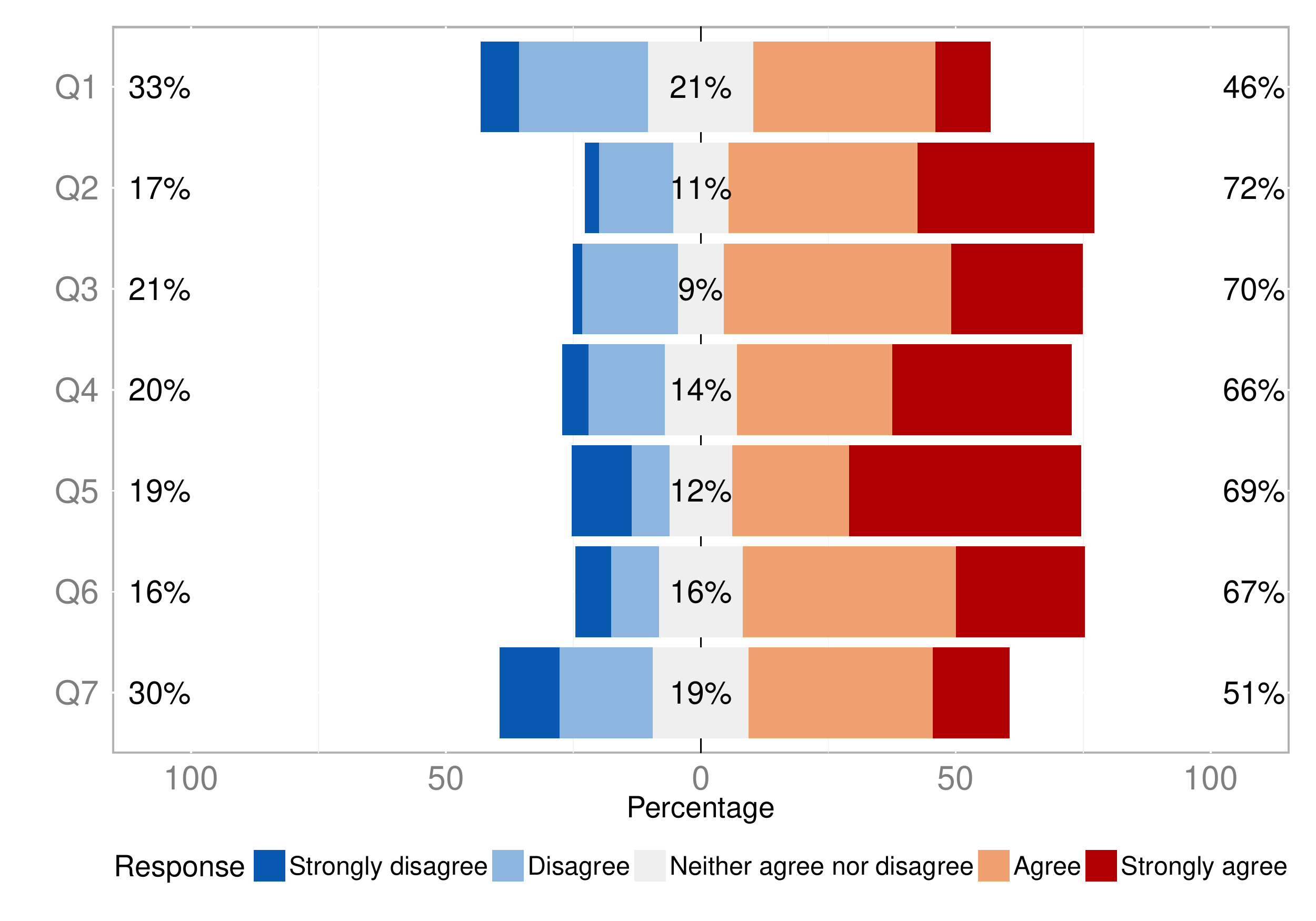}
	\caption{
    Answers to the post-test questionnaire.
    Items Q1--Q4 were answered by all participants.
    Items Q5--Q7 were answered only by participants in the experimental groups (B, C, and D).
	Percentages on the left side include participants that answered ``Strongly disagree'' or ``Disagree''.
	Percentages in the middle account for participants that answered ``Nor agree, nor disagree''.
	Percentages on the right side include participants that answered ``Agree'' or ``Strongly agree''.}
	\label{plot:compositelikert}
\end{figure}

We invited participants that completed all tasks to fill-in a post-test questionnaire.
Participants of all groups answered items related to their concerns about
malicious software on their smartphones and on the secrecy of their passwords,
both for email and e-banking.
Figure~\ref{plot:compositelikert} shows participant answers on 5-point Likert-scales.
33\% of the participants were not concerned,
21\% had neutral feelings, and 46\% were concerned about malicious software on
their smartphones (Q1). Most participants
reported that they were concerned about password secrecy (72\% for their e-banking passwords (Q2)
and 70\% for their email passwords (Q3)).
Participants were more concerned (66\% agrees or strongly agrees) about the secrecy of their
e-banking password than about the secrecy of their email passwords (Q4).

Participants in groups B, C, and D answered additional questions about personalized indicators.
Only 19\% of the participants were familiar with personalized indicator prior to this study.
69\% of the participants reported that they checked the presence of the indicator at each log in (Q5).
67\% of the participants found personalized indicators user-friendly (Q6), and
51\% of them would use personalized indicators in other application scenarios (Q7).
Appendix~\ref{app:posttestI} provides the full text of items Q1--Q7.

We asked participants of the experimental groups if they had noticed anything unusual when logging in to complete task 4.
To those participants that gave a positive answer we also asked if they logged in and
why. 23\% of the participants did not notice anything unusual, while 23\% did not remember.
54\% of the participants noticed something was wrong with the SecBANK
application while they were logging in. 36\% of them reported
that they logged in anyway. In the following section, we discuss the reasons
why users logged in despite noticing something unusual in the SecBANK login
screen.


\subsection{Discussion}
\label{sec:discussion}

Roughly half of the participants that used personalized indicators detected the phishing attack and did not enter their credentials to the phishing application.
While many participants still fell victim of the attack, our user study shows a significant improvement compared to previous studies in the context of websites~\cite{schechter07sp,lee-w2sp14}.
We conclude that personalized indicators empower alert users to detect phishing attacks in mobile applications.
Furthermore, the deployment of personalized indicators incurs no cost to the marketplace operator or to the OS developer, and incurs very little cost to the application provider.

The post-test questionnaire revealed that some participants entered their credentials even though they noticed that the indicator was missing.
We asked the reason for their behaviour and we report some answers (redacted to correct grammatical errors):

\begin{quote}
  ``Because it is a test and I had nothing to lose or hide.''
\end{quote}
\begin{quote}
  ``I knew this was a test product so there could not possibly be malware for it
  already. I just thought maybe you guys had some difficulties going on.''
\end{quote}

From the answers we could assume that the effectiveness of personalized indicators to detect phishing in a real-life scenario may be higher that the one we have reported.

Nevertheless, we also received answers such as:

\begin{quote}
  ``I thought it was a problem of the app [...] as it happens sometimes in Safari or other browsers.''
\end{quote}
\begin{quote}
  ``I thought it was a temporary bug''
\end{quote}

This shows that when it comes to IT-systems, users are expecting errors and bugs and therefore they are prone to fall for attacks that leverage the unreliable view that people have about computer products.

Finally, we discuss a few hypotheses to answer the question of what makes personalized indicators more effective in the context of mobile applications compared to that of
desktop environment. Mobile user interfaces are considerably simpler than the ones of desktop
websites. As the user focus is limited to a few visual
elements~\cite{rensink2002}, personalized indicators could be more salient in
mobile application UIs. The usage patterns of mobile applications may also be
different from those of websites. 

An interesting direction for future work is to study how to display personalized indicators to improve phishing detection.
For example, indicator position and size within the screen estate may change the effectiveness of indicators.
Evaluating different types of indicators (e.g., interactive indicators,
personalized UIs, etc.) is another direction for future work.


%% file: setup.tex
\section{Secure Setup of Application Indicators}
\label{sec:setup}

\begin{figure}[t]
    \centering
    \includegraphics[width=\linewidth]{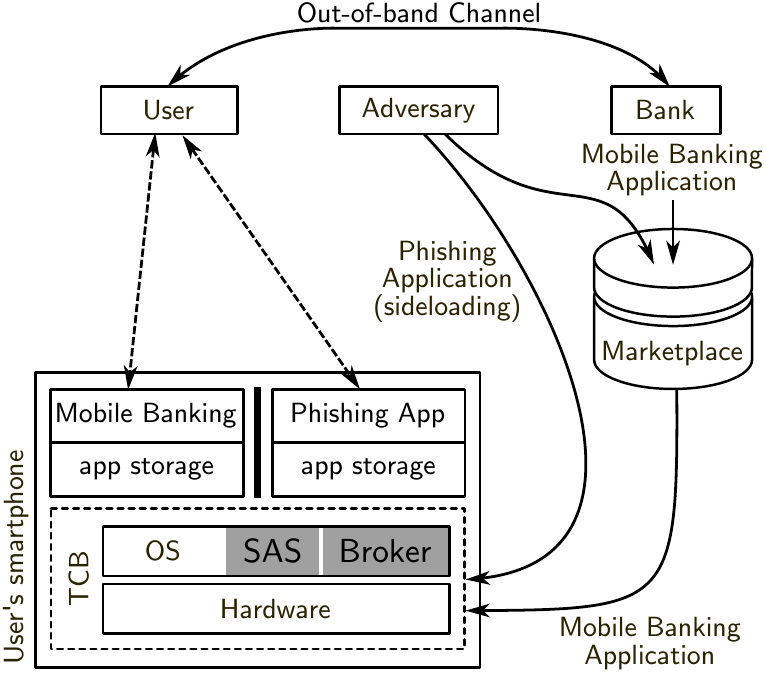}
    \caption{
    System model overview. Each application running on the smartphone is sandboxed and cannot access the memory or storage of other applications. 
    Applications are installed from the marketplace or via sideloading.
    The bank has an out-of-band channel (e.g., regular mail) to its customers.
    Our protocol adds two components to the platform TCB (shown in gray).
    The \secmodule{} can measure installed applications and has a UI to receive user inputs.
    A Secure Attention Sequence (SAS) is used to start the \secmodule{}.
    }
    \label{fig:system-model}
\end{figure}

So far we have presented evidence that personalized indicators can help users to detect application phishing attacks.
In this section we focus on the setup of personalized indicators and propose a secure protocol to bootstrap indicators in mobile platforms.

Setup of indicators in previous research proposals~\cite{tygar96woec,dhamija05soups,zu-woot12} and deployed systems~\cite{boa,vanguard} relies on the ``trust on first use'' (TOFU) assumption.
The indicator setup happens the first time the user registers for an online service~\cite{tygar96woec,dhamija05soups,boa,vanguard} or starts the application~\cite{zu-woot12},
assuming that the adversary is not present at this time.
Otherwise, if the indicator is phished during its setup, malicious software can later on masquerade as the legitimate application.

In the rest of this section we propose a protocol to setup indicators that does not rely on the TOFU assumption and, therefore, can withstand
phishing attacks when the user chooses the indicator.
As a use case, we consider a mobile banking scenario similar to the one of Section~\ref{sec:userstudy}, and consider an application that supports indicators to improve phishing detection.
We start by detailing the system and adversarial model we consider. 
Later on, we describe the indicator setup protocol and present an implementation for the Android platform.
Finally, we show the results of both performance and usability study.

\subsection{System and Adversary Model}
\label{sec:setup:system_and_adversary}
Figure~\ref{fig:system-model} illustrates the system model we consider. (Gray components are part of our solution and are detailed below.) Mobile applications run on top of an OS that, together with the underlying hardware, constitute the device TCB.
The TCB guarantees that a malicious application cannot tamper with the execution of another application (isolated execution) or read its data (application-specific storage).
The application active in foreground controls the screen and receives user inputs.
The TCB enforces that applications running in the background do not intercept user inputs intended for the foreground application and cannot capture its output (user interaction isolation).

Since we consider a mobile banking scenario, we assume the existence an out-of-band-channel between the bank (i.e., the application provider) and the user.
An example of this channel is the (regular) mail system routinely used by banks as a secure and authentic channel to transfer PINs and other credentials to their customers.
As currently used to reset forgotten passwords, emails can also be used as an alternative out-of-band channel.

The goal of the adversary is to phish the indicator that the user chooses for the banking application.
We assume that the user has previously installed, either from the marketplace or via sideloading, a malicious application on his smartphone.
Leveraging the malicious application on the user's phone, the adversary can launch any of the attacks presented in Section~\ref{subsec:attacks}.
However, the adversary cannot compromise the device TCB or the out-of-band channel between the bank and the user.

\subsection{Secure Indicator Setup Protocol}
\label{sec:setup:protocol}


Setting up an indicator in the presence of malicious software, requires the user to identify the banking application for which he is choosing the indicator.
A similarity attack to phish the indicator at setup time may be hard for the user to detect.
Similarly, the marketplace operator may fail to identify phishing applications and block their distribution (see Section~\ref{sec:comparison}).
We argue that no party but the bank can attest the validity of the banking application for which the user is about to choose the indicator.
For this reason, our protocol relies on a trusted component of the mobile platform that, together with the bank, attests the validity of the banking application installed on the device.

In particular, the trusted component and the bank establish an authentic channel to attest the application.
If attestation is successful, the trusted component provides the application with a PIN known by the user. 
The user can identify the legitimate application, if it shows the correct PIN.

Figure~\ref{fig:system-model} shows in gray the components that we add to the device TCB.
A system component that we call the \secmodule{} can measure applications (e.g., compute a hash of the application binary and its configuration files) installed on the device and has a UI to receive user inputs.
The mobile OS is also enhanced with a Secure Attention Sequence (SAS), which is a common approach to start a particular software component of a TCB~\cite{gligor87tse, mccune09ndss, libonati11ndss}\footnote{A popular SAS is the ``ctrl-alt-del'' sequence in Windows systems which generates a non-maskable interrupt that starts the user-logon process.}.
We implement the SAS operation as two repeated presses of the home button and we use it to start the \secmodule{}.
When the \secmodule{} is running, the mobile OS ensures that no background application can activate to the foreground.

The bank and the \secmodule{} establish an authentic channel using the out-of-band channel between the bank and the user.
The bank sends a measurement of the legitimate banking application over the authentic channel, so that the \secmodule{} can compare it with the measurement of the banking application installed on
the device. If the two measurements match, the \secmodule{} transfers to the banking application a PIN known by the user.
The banking application shows the PIN to the user who can, therefore, identify the legitimate banking application.

\begin{figure}[t]
    \centering
    \includegraphics[width=\linewidth]{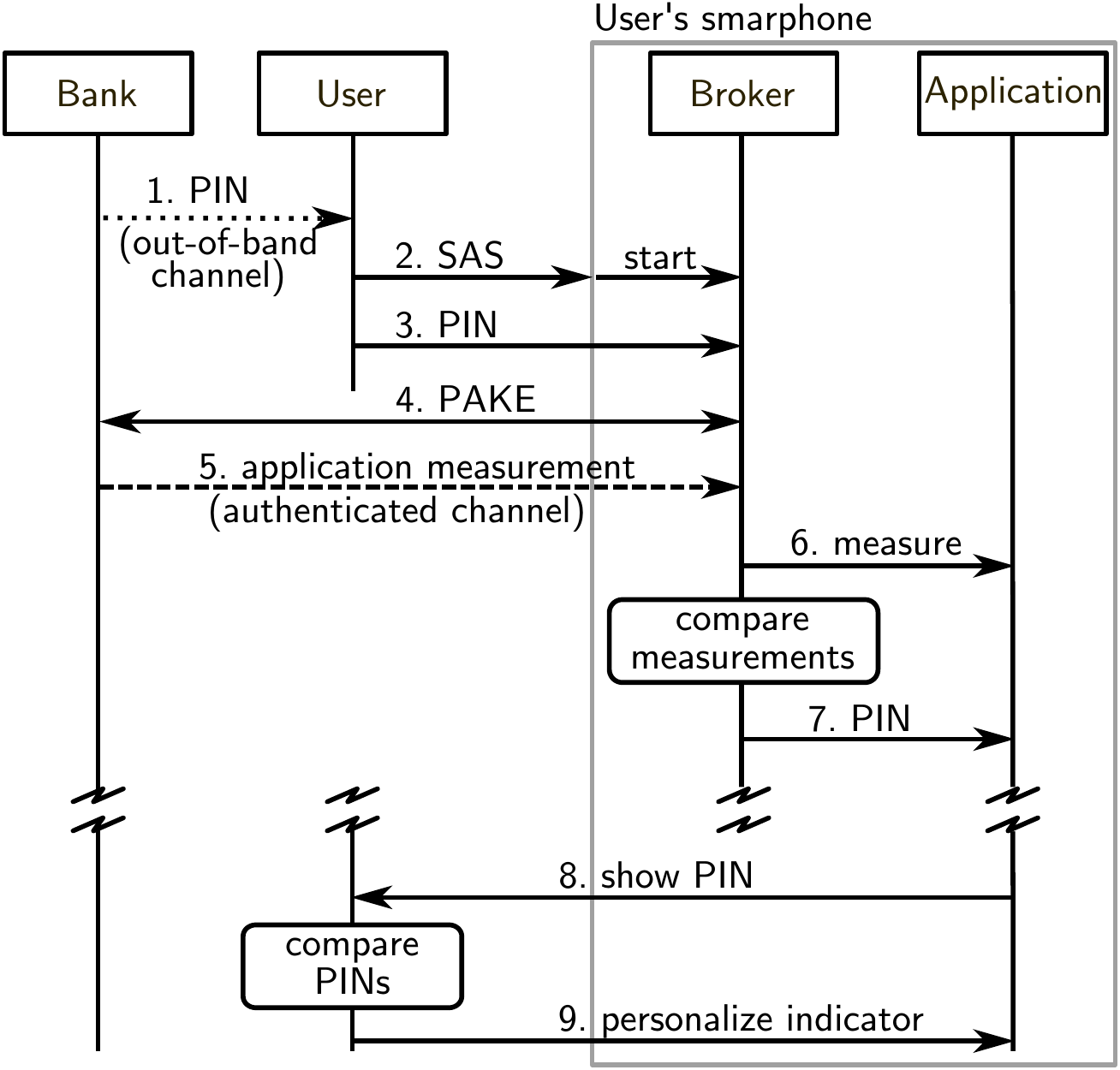}
    \caption{The personalized indicator setup protocol. The user performs an SAS operation to start the \secmodule{} and enters the PIN.
    The \secmodule{} verifies the installed banking application with the help of the bank.
    If the installed application is the legitimate one, the \secmodule{} starts it and the user can choose a custom indicator.}
    \label{fig:protocol}
\end{figure}

Figure~\ref{fig:protocol} illustrates the steps to securely setup a personalized indicator. Here we explain them in detail.

\begin{enumerate}
\item The bank uses the out-of-band channel (e.g., regular mail) to send a PIN to the user.

\item The user installs the application, downloading it from the marketplace.
    When the installation completes, the user performs the SAS operation to start the \secmodule{}.
    While the \secmodule{} is running, the mobile OS prevents third-party applications from activating to the foreground.

\item   The user inputs the PIN to the \secmodule{}.

\item   The \secmodule{} and the bank use the PIN to run a Password Authenticated Key Exchange (PAKE)
        protocol~\cite{speke} and establish a shared key.

\item   The bank sends the measurement of the legitimate banking application to the \secmodule{}.
        The measurement can be the hash of the application installation package.
        The message is authenticated using the key established during the previous step.
	
\item   The \secmodule{} verifies the authenticity of the received message, measures the banking application on the device,
        and checks its measurement against the reference value received from the bank.

\item   If the two measurements are identical, the \secmodule{} securely transfers the PIN to the banking application
        (e.g., writes the PIN to the application-specific storage).
        Otherwise the \secmodule{} aborts the protocol and notifies the user.

\item   The \secmodule{} starts the banking application and the mobile OS restores the functionality that allows background applications to
        activate to the foreground. The banking application displays the PIN which serves
        as a confirmation to the user that the application in foreground is the legitimate banking application.

\item   The user identifies the application in foreground as the legitimate banking application if it displays the same PIN that the user has received from the bank.
        At this point, the user can choose a personalized indicator for the banking application.
\end{enumerate}

\paragraph{Security Analysis.}

Our protocol relies on the user attention when setting up the indicator.
At the beginning of the setup protocol, the user must input the PIN only after performing the SAS operation.
At the end of the protocol, the user must choose an indicator only if the application in foreground displays the same PIN that the user has received from the bank.

The SAS operation and the device TCB guarantee that no information is leaked when the user inputs the PIN to the \secmodule{}.
In particular, no malicious application can masquerade as the \secmodule{} to phish the PIN.
The PIN is used as a one-time password to derive a key through the PAKE protocol.
The derived key is only used to authenticate one message (from the bank to the \secmodule{}).
The security properties of the PAKE protocol guarantee that, given a transcript of the protocol, the adversary can neither learn the PIN, nor brute-force the set of possible PINs~\cite{spekesec}.

The application-specific storage functionality, provided by the mobile OS, guarantees that the PIN received by the banking application can not be read by
other applications.

A phishing application on the device will not receive the PIN from the \secmodule{} because its measurement differs from the one of the legitimate banking application.
The adversary cannot impersonate the bank to the \secmodule{} without knowledge of either the PIN or the key derived through the PAKE protocol.

\subsection{Implementation}
\label{sec:setup:implementation}


We build a prototype of our setup protocol for the Android platform.
We use a Samsung Galaxy S3 and develop against the CyanogenMod Android OS (version 4.3 JellyBean).
Cryptographic operations are based on the Bouncy Castle library~\cite{bouncycastle}.
Message authentication uses HMAC-SHA256 with a 128-bit key.
The bank's server is implemented in Python, using the CherryPy Web Framework and SQLite.
Client-server communication works over a standard HTTP channel using messages in the JSON standard format.
Our implementation adds two Java files, to implement the \secmodule{} as a privileged application, and modifies four system Java files.
A total of 652 lines of code were added to the system TCB.

We add an extra tag (\texttt{<secureapk>}) to the Android Manifest file of the banking application.
The tag contains two fields indicating a URL and an application handle.
The \secmodule{} uses the URL to contact the application provider (i.e., the bank) and the handle to identify the application to attest.
When the banking application is installed, the \texttt{PackageParser} of the Android OS reads the extra information in the \texttt{<secreapk>} tag and stores it for later use.

\begin{figure}[t]
  \centering
  \subfloat[] {%
    \includegraphics[width=.45\linewidth]{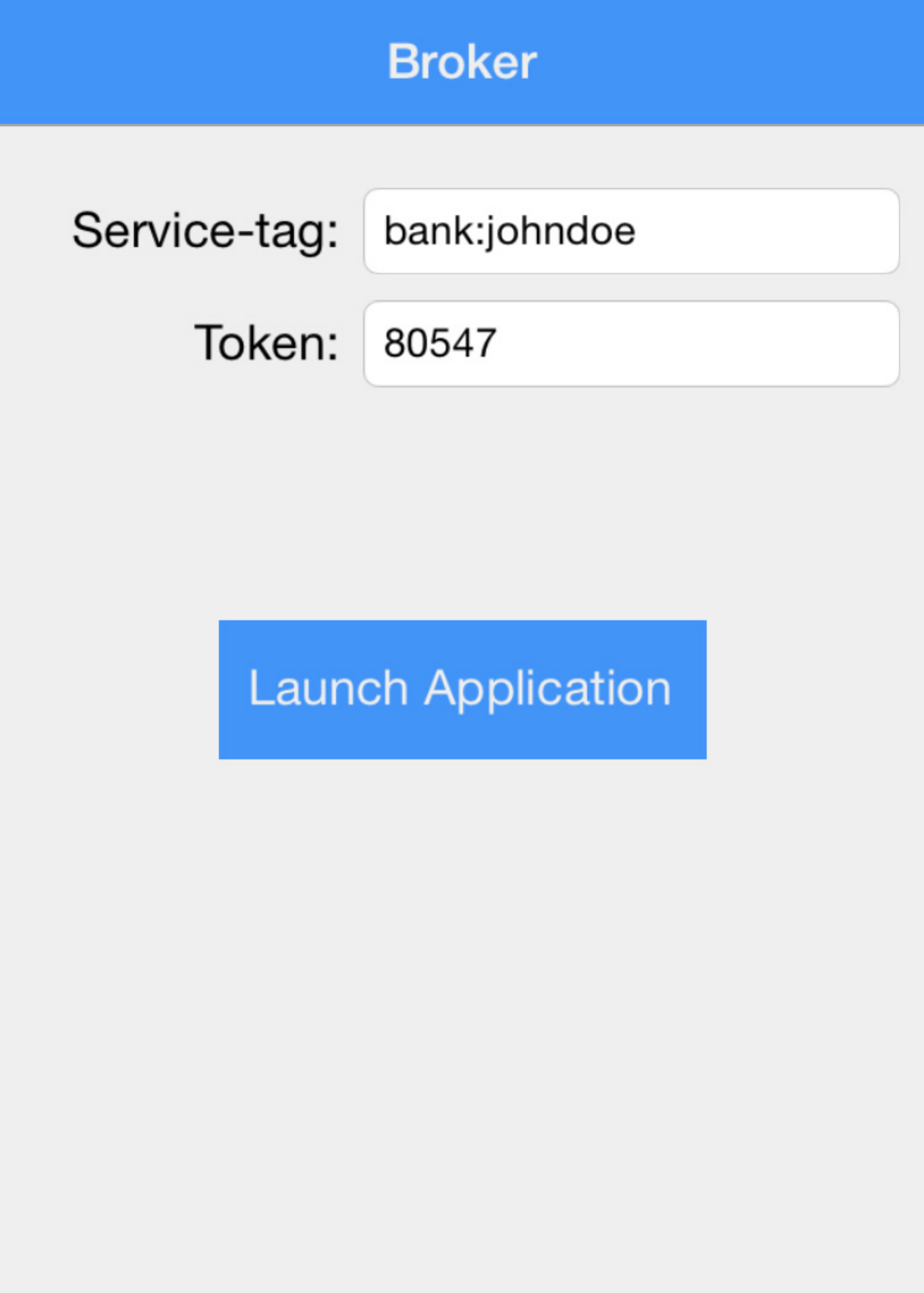}
    \label{fig:broker}
  }
  ~
  \qquad
  \subfloat[]{%
    \includegraphics[width=.45\linewidth]{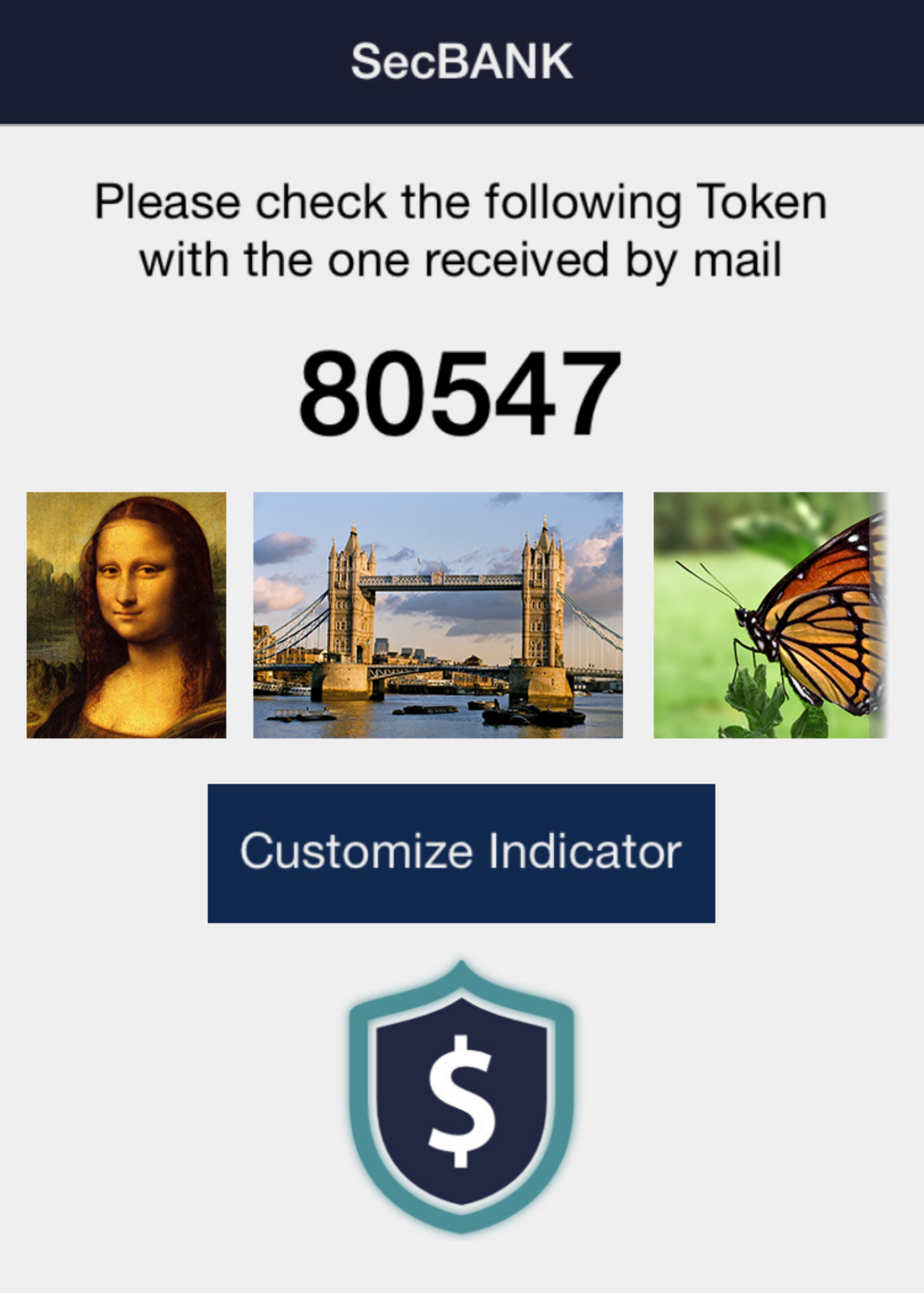}
    \label{fig:tokenvoid}
  }
  \caption{
  (a) Screenshot of the \secmodule{} prototype. The user inputs the service tag and the PIN received by the bank.
  (b) Banking application started by the \secmodule{}. The user must verify that the PIN displayed matches the one received by the bank.
  The application shows the images in the local photogallery and the user can select one as the indicator.}
\end{figure}

We assign the SAS operation to a double-tap on the home button.
The SAS operation unconditionally starts the \secmodule{} that asks the user to enter the PIN received from the bank (see Figure~\ref{fig:broker}).
In our implementation, the bank sends to the user a 5-digits PIN (e.g., ``80547'')
and a \emph{service tag}. The service tag contains an application handle used by the \secmodule{}
to search through the registered handles and to identify the application to attest.
The service tag also contains a user ID, sent to the bank by the \secmodule{}, to retrieve the correct PIN from its database.
An example of a service tag is ``bank:johndoe'', where ``bank'' is the application handle and ``johndoe'' is the user ID.
After the user has input the service tag and the PIN, the \secmodule{} uses the handle to identify the application to attest.
At this time the \secmodule{} also fetches the URL stored by the \texttt{PackageParser}, to contact the bank's server.

\begin{table}[t]
  {\small
  \begin{center}
    {\tabulinesep=.6mm
      \setlength{\tabcolsep}{1.5mm}
      \begin{tabu}{|l|l|} \hline
        TCB increment           & 652 LoC \\ \hline
        Communication overhead  & 705 bytes \\ \hline
        Execution time (WiFi)   & 421ms ($\pm$21ms)\\
        Execution time (3G)     & 2042ms ($\pm$234ms)\\
        Execution time (Edge)   & 2957ms ($\pm$303ms)\\ \hline
      \end{tabu}}
  \end{center}}
  \caption{Evaluation summary for the personalized indicator setup prototype.}
  \label{tab:implementation}
\end{table}

We use SPEKE~\cite{speke} as an instantiation of the key establishment protocol.
SPEKE uses 4 messages, including a key confirmation step.
The first SPEKE message sent by the \secmodule{} also contains the user ID that allows the bank's server to select the correct PIN from its database.
The server uses the key established via the PAKE protocol to compute a MAC over the hash of the legitimate application.
Both the hash and the authentication tag are sent to the \secmodule{}.
The \secmodule{} verifies the authentication tag, hashes the installed application, and compares the hash computed locally with the
received one. If the two hashes match, the \secmodule{} writes the PIN to the application's folder so that it can only be read by
that application. Otherwise, the \secmodule{} aborts and notifies the user.

Figure~\ref{fig:tokenvoid} shows the banking application started by the \secmodule{}.
The user is asked to compare the displayed PIN with the one received via mail and choose a personalized indicator.

\paragraph{Evaluation.}

We evaluate the setup protocol using a a sample banking application of 264KB.
Client-server communication overhead totals to 705 bytes, of which 641 bytes account for the SPEKE protocol. (Communication overhead is independent of the size of the application.)
We test the protocol over a WiFi connection (hosting the server in a remote location, i.e., not on a local network), as well as over 3G and EDGE cellular data connections.
Table~\ref{tab:implementation} summarizes the evaluation in terms of added lines of code, communication overhead, and execution time.
We note that hashing the banking application takes  25ms on average, with a standard deviation of 2ms.
The time to hash an application increases linearly with the size of the application and remains lower than the network delay for applications up to 100MB.

\subsection{Usability}
\label{sec:setup:usability}
We run a small-scale user study to evaluate the usability of our setup protocol.
In particular, our goal was to understand how easy it is for users to setup indicators by following written instructions, like the ones a bank customer would receive from his bank.
We also wanted to verify how attentive users are when comparing the PIN shown by the application with the one received from the bank.

We invited participants to our lab and asked them to carry out the set up protocol as if they were customers of a bank that uses indicators in its application.
We provided participants with a phone and a letter from the bank with instructions on how to setup the indicator.
We assigned participants to three different groups.
One group carried out the set up protocol in benign settings.
For the remaining two groups, we simulated variants of a background attack at the time when the banking application displays the PIN (step 8 of the setup protocol).
We recorded whether participants pressed the ``Customize Indicator'' button and chose a personalized indicator while under attack.

\paragraph{Procedure.}
\emph{Recruitment and Group Assignment.}
We advertised the study as a user study ``to evaluate the usability of a setup protocol for an e-banking application''.
Participants were asked to come to our lab and setup the application on a smartphone that we provided.
We informed participants that we would not collect any personal information and offered a compensation of 20\$.

We selected 30 participants and assigned them to one of three groups in a round-robin fashion. (None of these participants were involved in the user study of Section~\ref{sec:userstudy}.)
The difference among the groups was the PIN shown by the banking application right before participants were asked to choose a personalized indicator.
Group A participants were shown the same PIN that appeared  on the letter from the bank.
Group B participants were shown a random PIN.
Group C participants where shown no PIN.

Participants of group A reflected the user experience of the setup protocol under no attack.
Participants of groups B and C reflected the user experience of the setup protocol under a background attack.

\emph{Experiment.}
The experiment started with a pre-test questionnaire to collect demographic information.
After the questionnaire, participants were given a smartphone and  a letter from a fictitious bank with instructions to setup the indicator.
The letter included a 5-digit PIN and a service tag to be used during the setup protocol.
The procedure was the one described in Section~\ref{sec:setup:protocol} and was carried out on a Galaxy S3.
We stress that we did not interact with the participants for the duration of the setup protocol.
Participants were also asked to fill in a post-test questionnaire to evaluate the procedure they had just completed.

\paragraph{Results.}

37\% participants were males and 63\% were females. Most of them had completed a university degree (73\%), and were aged between 24 and 35 (80\%).

All participants in group A managed to complete the task successfully.
All participants of group B aborted the setup procedure because they detected a mismatch between the PIN displayed by the banking application and the one in the letter from the bank.
40\% percent of the participants in group C failed to detect the missing PIN and completed the setup process, thereby leaking the indicator to the phishing application.

\begin{figure}[t]
	\centering
	\includegraphics[width=\columnwidth]{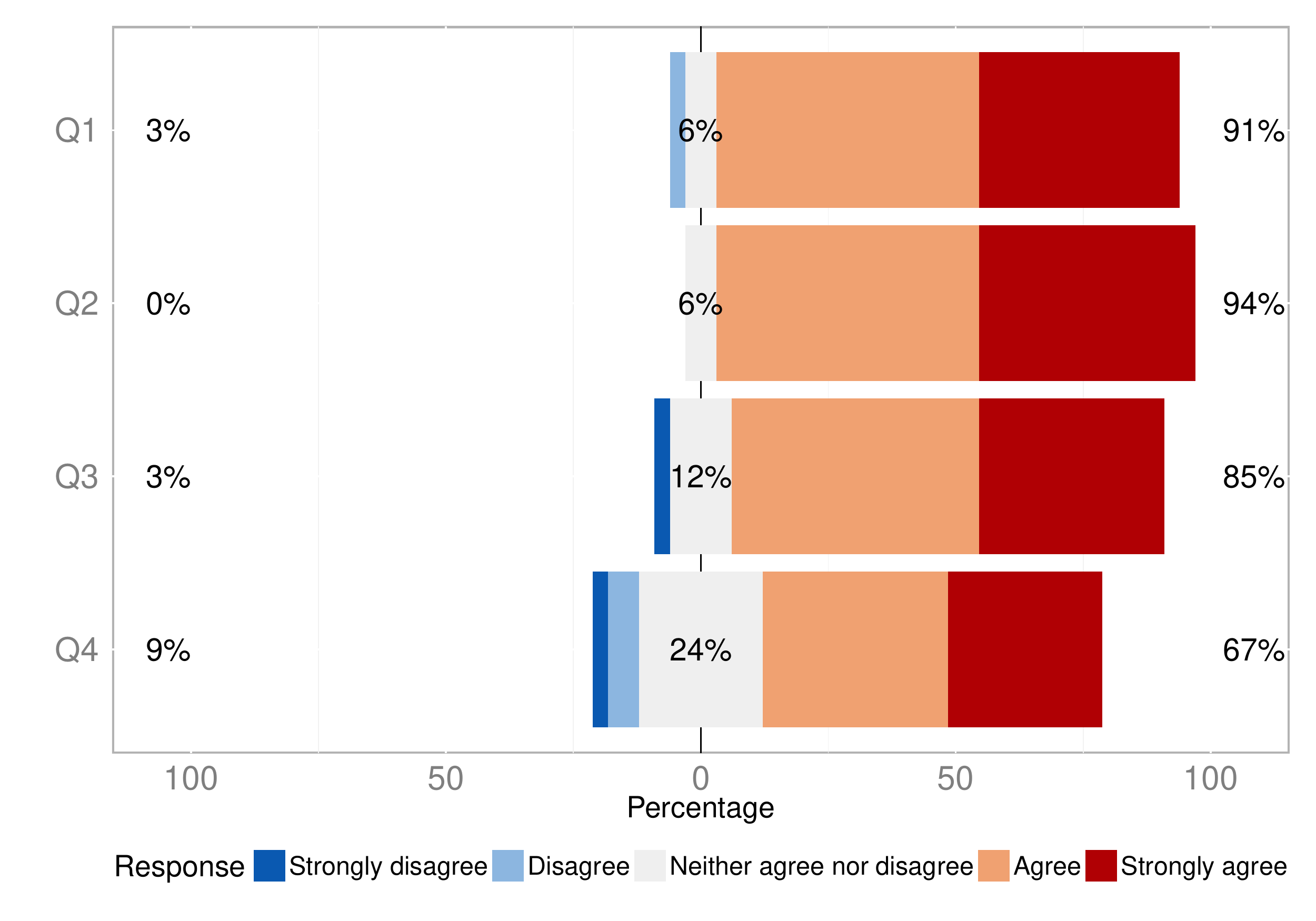}
	\caption{
    Answers to the post-test questionnaire for the user study on the indicator setup protocol.
	Percentages on the left side include participants that answered ``Strongly disagree'' or ``Disagree''.
	Percentages in the middle account for participants that answered ``Nor agree, nor disagree''.
	Percentages on the right side include participants that answered ``Agree'' or ``Strongly agree''.}
	\label{plot:second_likert}
\end{figure}

The post-test questionnaire revealed that 91\% of the participants rated the instruction sheet easy to understand (Q1) and 94\% of them rated the task easy to complete (Q2).
85\% of the participants believed that they had completed the task successfully (Q3) and 67\%  of them would use the mechanism in a mobile banking application (Q4).
Figure~\ref{plot:second_likert} shows the distribution of the answers.
Appendix~\ref{app:posttestII} provides the full text of items Q1--Q4.

Our study suggests that the setup procedure was simple enough to be completed by average users.
Regarding attack detection, the ``missing PIN'' attack was the only one that went undetected.
Participants may have judged the absence of the PIN as a temporary bug and decided to continue with the setup protocol.
Similarly to what we have reported in Section~\ref{sec:discussion}, this result shows that users are acquainted with software bugs and
are likely to tolerate a temporary bug even during a security-sensitive operation.
Nevertheless, the user study reveals that our solution is usable and effective.


%% file: relatedwork.tex

\section{Related Work}
\label{sec:related}

\noindent\textbf{Application phishing.}
Application phishing attacks have been described in recent research~\cite{chin11mobisys, zu-woot12, felt11w2sp} and several attacks have been
reported in the wild~\cite{droid09, digitaltrends, securelist}.
Proposed solutions are primarily ad-hoc, i.e., they identify specific
attacks and try to prevent them by restricting the device functionality that
those attacks exploit~\cite{zu-woot12, chin11mobisys, hou-umich12}.
Personalized security indicators are a well-known concept from the context of website phishing and can
provide an holistic solution to application phishing attacks in mobile platforms.
Similar to previous work on website phishing, the authors of~\cite{zu-woot12} assume that the adversary
cannot mount a phishing attack when the indicator is set up.
In this paper we described a protocol that allows the secure setup of personalized indicators
in presence of an adversary.
We are not aware of any previous work that provides the same security property.

\noindent\textbf{Web phishing.}
Anti-phishing techniques for the web have been deployed in practice~\cite{dhamija06chi, hong-cacm12}.
Proposals include automated comparison of website URLs~\cite{maurer-css12},
visual comparison of website contents~\cite{chen10tit, yue-www07},
use of a separate and trusted authentication device~\cite{parno-fincrypto06},
personalized indicators~\cite{dhamija05soups,schechter07sp,lee-w2sp14},
multi-stage authentication~\cite{herzberg12},
and attention key sequences to trigger security checks on websites~\cite{wu-soups06}.
While some of these mechanisms are specific to the web environment, others could be adapted also for mobile
application phishing detection. Website phishing in the context of mobile web
browsers has been studied in~\cite{niu-upsec08,rydstedt-woot10}.

\noindent\textbf{Effectiveness of Security Indicators}
Previous research on the effectiveness of security indicators has mostly focused on phishing on the web.
Studies in this context have shown that users tend to ignore security indicators such as
personalized images~\cite{schechter07sp,lee-w2sp14} or toolbars~\cite{wu-chi06, jackson-fincrypto07}.
Browser warnings have been evaluated positively in a recent work~\cite{akhawe13usenix}, even if previous
studies suggest otherwise~\cite{sunshine-usenix09,egelman-chi08,dhamija06chi}.

\noindent\textbf{Mobile malware.}
Malicious mobile applications typically exploit platform vulnerabilities (e.g.,
for privilege escalation) or use system APIs that provide access to
sensitive user data.
A malicious application can, for example, leak the user location or send SMS messages to premium numbers without user consent.
For reviews on mobile malware, we refer the reader to recent surveys~\cite{zhou-jiang12sp, felt11spsm}.
Infection rates of mobile malware are reported in~\cite{truong13}.

Mobile malware detection is typically based on the analysis of system call patterns and detection of known platform exploits.
The authors of~\cite{lever-ndss13} propose to detect malware by analyzing network traffic patterns.
The detection mechanisms used for mobile malware
are ill-suited for detection of mobile phishing applications, as many phishing
attacks do not exploit any platform vulnerability, do not require access to
security-sensitive system APIs, and do not generate predictable network traffic
patterns.


%% file: conclusion.tex
\section{Conclusion}
\label{sec:conclusion}

We have shown that personalized indicators can help users to detect application phishing attacks in mobile platforms and have very little deployment cost.
We have conducted a large-scale (221 participants) user study and have found that a significant number of participants that used personalized security indicators were able to detect phishing.
All participants that did not use indicators could not detect the attack and entered their credentials to a phishing application.
Based on these results we conclude that personalized indicators can help phishing detection in mobile applications and their reputation as an anti-phishing mechanism should be reconsidered.

We have also proposed a novel protocol to setup personalized security indicators that does not rely on the ``trust of first use assumption''.
Our protocol allows users to securely setup the indicator, despite phishing software on the device.
We have reported details on the implementation of the setup protocol and have evaluated its usability.

%% file: appendix.tex

\appendix

\section{Post-test Questionnaire I}
\label{app:posttestI}
We report the items of the post-test questionnaire for the user study on the effectiveness of personalized indicators (Section~\ref{sec:userstudy}).
Items Q1--Q4 were answered by all participants.
Items Q5--Q7 were answered only by participants in the experimental groups (B, C, and D).
All items were answered with a 5-point Likert-scale from \emph{Strongly Disagree} to \emph{Strongly Agree.}

\begin{itemize}
\item[Q1]\emph{I am concerned about malware on my smartphone}
\item[Q2]\emph{I am concerned about the secrecy of my password for mobile banking}
\item[Q3]\emph{I am concerned about the secrecy of my password for my email account}
\item[Q4]\emph{I am more concerned about the secrecy of my mobile banking password compared to my email account password}
\item[Q5]\emph{During the study, I have made sure that my personal image was showing before entering my username and password}
\item[Q6]\emph{The login mechanism using personal images was intuitive and user-friendly}
\item[Q7]\emph{I would use personal images in other applications}
\end{itemize}

%
%

\section{Post-test Questionnaire II}
\label{app:posttestII}
We report the items of the post-test questionnaire for the user study on the indicator setup protocol (Section~\ref{sec:setup}).
All items were answered with a 5-point Likert-scale from \emph{Strongly Disagree} to \emph{Strongly Agree.}

\begin{itemize}
\item[Q1]\emph{The instruction sheet was clear and easy to understand}
\item[Q2]\emph{The task was easy to complete}
\item[Q3]\emph{I believe that I have successfully completed the task}
\item[Q4]\emph{I would use this mechanism to improve the security of my mobile banking application}
\end{itemize}
